\documentclass[aps,prb,floatfix,twocolumn,a4paper,10pt,showpacs,
citeautoscript,superscriptaddress]{revtex4-1} 

\usepackage[english]{babel}	
\usepackage{amsmath}		
\usepackage{amssymb}		
\usepackage{bbm}			
\usepackage{graphicx}	
\usepackage{graphics}	
\usepackage{xcolor}		
\DeclareGraphicsExtensions{.pdf}
\usepackage[colorlinks=true, pdfstartview=FitV,linkcolor=blue, citecolor=blue, urlcolor=blue]{hyperref}

\makeatletter
\newcommand*{\Equation}{\@ifstar\sEquation\oEquation}
\newcommand{\sEquation}[1]{\begin{equation*}#1\end{equation*}}
\newcommand{\oEquation}[2]{  \begin{equation}\label{#1}#2\end{equation} }
\makeatother
\newcommand{\Align}[2]{\begin{align}\label{#1}#2\end{align}}
\newcommand{\SubAlign}[2]{\begin{subequations}\label{#1}\begin{align}#2\end{align}\end{subequations}}

\newcommand{\bs}{\boldsymbol}

\newcommand{\Partref}[1]{Section~\ref{#1}}
\newcommand{\Appref}[1]{Appendix~\ref{#1}}
\newcommand{\Figref}[1]{Fig.~\ref{#1}}
\newcommand{\Eqref}[1]{\eqref{#1}}

\newcommand{\eg}{{\it e.g.~}}

\newcommand{\resp}{{\it resp.~}} 

\newcommand{\Relative}{\mathbbm{Z}}
\newcommand{\Rational}{\mathbbm{Q}}
\newcommand{\Real}{\mathbbm{R}}

\newcommand{\groupU}[1]{\mathrm{U}(#1)} 
\newcommand{\groupCP}[1]{{\mathbbm{C}{P}}^{#1}} 
\newcommand{\CPone}{$\groupCP{1}\ $}
\newcommand{\Uone}{$\groupU{1}\ $}
\newcommand{\groupUU}{$\groupU{1}\times\groupU{1}\ $}

\newcommand{\Exp}[1]{\text{e}^{#1}}

\renewcommand\Im{\mathrm{Im}}

\newcommand{\Grad}{{\bs \nabla}}

\newcommand{\Curl}{\Grad\times}

\newcommand{\x}{{\bf x}}

\newcommand{\Et}{{\bf e}_\theta}

\newcommand{\F}{\mathcal{F}}

\newcommand{\oo}{{(1)}}
\newcommand{\ot}{{(2)}}
\newcommand{\oa}{{(a)}}

\newcommand{\D}{{\bf D}}
\newcommand{\A}{{\bf A}}
\newcommand{\B}{{\bf B}}
\newcommand{\J}{{\bf J}}

\newcommand{\Q}{\mathcal{Q}}

\newcommand{\ea}{e_a}

\newcommand{\ga}{g_a}
\newcommand{\gb}{g_b}

\newcommand{\psia}{\psi_a}
\newcommand{\psib}{\psi_b}
\newcommand{\varphia}{\varphi_{a}}

\newcommand{\varphiot}{\varphi_{12}}

\begin{document}

\title{Topological defects in mixtures of superconducting condensates with different charges}

\author{Julien~Garaud}
\email{garaud.phys@gmail.com}
\affiliation{Department of Physics, University of Massachusetts Amherst, MA 01003 USA }
\affiliation{Department of Theoretical Physics, Royal Institute of Technology, 
Stockholm, SE-10691 Sweden}
\author{Egor~Babaev}
\affiliation{Department of Theoretical Physics, Royal Institute of Technology, 
Stockholm, SE-10691 Sweden}
\affiliation{Department of Physics, University of Massachusetts Amherst, MA 01003 USA }

\date{\today}

\begin{abstract}

We investigate the topological defects in phenomenological models 
describing mixtures of charged condensates with commensurate electric 
charges. Such situations are expected to appear for example in liquid 
metallic deuterium. This is modeled by a multicomponent Ginzburg-Landau 
theory where the condensates are coupled to the same gauge field by 
different coupling constants whose ratio is a rational number. We 
also briefly discuss the case where electric charges are incommensurate.
Flux quantization and finiteness of the energy per unit length dictate 
that the different condensates have different winding and thus different 
number of (fractional) vortices. Competing attractive and repulsive 
interactions lead to molecule-like bound state between fractional 
vortices. Such bound states have finite energy and carry integer flux 
quanta. These can be characterized by $\mathbbm{C}P^1$ topological 
invariant that motivates their denomination as skyrmions.

\end{abstract}
\pacs{67.85.Jk,74.25.Ha, 67.85.Fg}

\maketitle

\section*{Introduction}

Although recently there has been substantial interest in multicomponent 
superconductors, typically the research is restricted to fields which 
have same value of electric charge modulus. See, e.g., 
Ref.~\onlinecite{Herland.Babaev.ea:10} for recent work with a field 
overview. In the typical condensed matter systems the charge is set by a 
Cooper pair charge of $2e$ for electronic systems or $-2e$ for protonic 
superconductors \cite{Babaev.Sudbo.ea:04,Babaev.Ashcroft:07}. By contrast, 
multicomponent systems with different values of electric charge attracted 
much less attention, yet some were discussed in the literature. One example 
is liquid metallic deuterium where deuteron is a charge-$1$ boson which 
can Bose condense and coexist with the Cooper pairs of electrons and/or 
protons \cite{Oliva.Ashcroft:84,Oliva.Ashcroft:84a,Babaev.Sudbo.ea:04,cherman}. 
Cooper pairs carry twice the charge of their constituent fermion, while a 
Bose-Einstein condensate of deuterons carries only once the charge of its 
(boson) constituent. This system is currently a subject of vigorous 
experimental pursuit \cite{eremets}.

Mixtures of condensates carrying different electric charges may also 
apply to ultra-cold atomic gases with synthetic gauge field. Recent 
progress both in theoretical understanding and experimental techniques 
to control such systems makes it promising that artificial dynamical 
gauge fields may be realized there \cite{Banerjee.Dalmonte.ea:12,
Zohar.Cirac.ea:12}. In that case, effectively a system could be described 
by multicomponent charged condensate models. Our present discussion may, 
in the future, find applications to these systems.

In superconductors with more than two components and $U(1)^N$ broken
symmetry there can be pairing transitions to $U(1)^{N-M}$ paired 
states driven by proliferation of composite vortices \cite{npba,kuklov,
Smorgrav.Babaev.ea:05}. Such pairing mechanisms can lead to charge-$4e$ 
electronic superconducting systems as hypothesized recently in various 
contexts \cite{agterberg,berg,Herland.Babaev.ea:10}, along with other 
recently discussed mechanisms for charge-$4e$ superconductivity \cite{moon}.

The above examples from superconductivity and superfluidity, along with 
the multicomponent gauge theories which appear as effective field theories 
in other condensed matter systems \cite{Sachdev:08,Chen.Huang.ea:13}, 
calls for investigation of mixtures of charged condensates with arbitrary 
ratio of condensate charges. To this end we study a phenomenological 
Ginzburg-Landau model that accounts for such mixtures, regardless of 
their underlying microscopic origin.
We discuss below that if the mixtures of charged condensates carrying 
different electric charges are realized, either in natural or artificial 
systems, their response to external applied magnetic field or rotation 
would be very different from those of mixtures with equal charges. 
This is because of the substantial difference in the topological 
excitations as compared to systems where condensates carry the 
same electric charge.


In \Partref{Model} we introduce a mean-field model that accounts 
for mixtures of charged condensates, when the condensates carry 
different electric charges. \Partref{Fractional} is devoted to 
elementary topological excitations, that is fractional vortices, 
and flux quantization in our Ginzburg-Landau model. In \Partref{LL}, 
we investigate the physics of flux carrying topological excitations 
within the London limit where condensates are assumed to have constant 
densities.

\section{Mixtures of charged condensates} 
\label{Model}

A charged condensate can, under certain conditions, be described by 
mean field Ginzburg-Landau free energy that couples it to the vector 
potential of the magnetic field through the kinetic term 
\Equation{}{
\frac{1}{8\pi}|\Curl\A|^2 +\frac{\hbar^2}{2m}
\left|\left(\Grad+i\frac{e^*}{\hbar c}\A\right)
\psi\right|^2
+V(\psi) \,.
}
Here $e^*$ is the electric charge of the condensate and $m$ its rest 
mass. $\hbar$ and $c$ are respectively the reduced Planck constant 
and the speed of light in the vacuum. $V$ is the interacting potential.
For example, for an ordinary superconductor, a Cooper pair has charge 
twice that of an electron. Thu, there $e^*=2e$. On the other hand, 
a Bose-Einstein condensate of singly charged bosons will have electric 
charge $e^*=e$. Thus mixtures of charged condensates, should generically 
have different couplings to the vector potential. A mixture of condensates 
with different masses and charges is thus described by 
\Equation{}{
\frac{1}{8\pi}|\Curl\A|^2 +\sum_a\frac{\hbar^2}{2m_a}
\left|\left(\Grad+i\frac{e_a}{\hbar c}\A\right)\psi_a\right|^2
+V(\psi_a)
\,.
}
Note that since the theory is invariant under complex conjugation, 
it is sufficient to consider positive charge only, without losing 
generality, condensates with negative charge being obtained by 
complex conjugation of the one with the positive charge. Now, to 
get rid of superfluous parameters, we express the energy in 
units of $\frac{\hbar^2c^2}{4\pi}$ and rescale the fields as
\Equation{Rescale}{
\tilde{\A}=\frac{\A}{\hbar c}\,,~~~\text{and}~~~
\tilde{\psi}_a=\sqrt{\frac{4\pi}{m_ac^2}}\psi_a\,.
}
Dropping the $\tilde{~}$ symbol further on, a mixture of charged 
condensates is thus described by the free energy density 
\Equation{FreeEnergy}{
 \F= \frac{1}{2}(\Curl\A)^2  
   +\sum_{a=1,2}\frac{1}{2}|(\Grad+i\ea\A)\psi_a|^2   
	+V(\psi_a) \,.
}
Here $\psi_a=|\psi_a| e^{i\varphi_a}$ are complex fields that  
stand for the charged condensates (with different indices $a=1,2$).
The condensates are coupled together through the electromagnetic 
interactions mediated by the vector potential $\A$ in the kinetic 
terms $\D\psi_a=(\Grad+i\ea\A)\psi_a$. Since $\psi_a$ are condensates 
that are essentially different, they should be independently conserved. 
It results that the potential has global \groupUU invariance 
ensuring independent conservation of both particle numbers
\Equation{Potential}{
V(\psi_a)=\sum_a\alpha_a|\psi_a|^2+\frac{1}{2}\beta_a|\psi_a|^4 \,.
}
In the condensed phase, $\alpha_a$ are negative parameters while  
$\beta_a>0$. The model exhibits gauge invariance under local \Uone 
transformations. That is, for arbitrary $\chi(\x)$, the energy 
\Eqref{FreeEnergy} is unchanged under the transformations 
\Equation{GaugeTR}{
\A\rightarrow\A-\Grad\chi\,,~~~
\psi_a\rightarrow\Exp{ie_a\chi}\psi_a\,.
}

As will be explained later on, a necessary condition for finite 
energy flux carrying configurations is that the charge of the 
condensates should be commensurate. That is, the ratio of the 
coupling constants $\ea$ is a rational number:
$e_1/e_2\in\Rational$. To capture this constraint, it is convenient 
to parametrize the gauge couplings as $\ea=e\ga$ where $\ga$ are 
integer numbers ($\ga\in\Relative$). Moreover, the two integers 
$g_1$ and $g_2$ should be relatively prime (their greatest common 
divisor must be $1$). Within our parametrization, $e$ is an arbitrary 
number that uniquely parametrizes the London penetration length 
defined below.
If we apply this model to liquid metallic deuterium, the couplings 
are $(g_1,g_2)=(1,2)$ and $\psi_1$ denotes the deuteron condensate 
while $\psi_2$ (carrying twice the electric charge of $\psi_1$) 
denotes electronic Cooper pairs.

Functional variation of the free energy \Eqref{FreeEnergy} 
determines the Euler-Lagrange equations of motion. That is, 
variation with respect to complex fields $\psi_a^*$ gives 
the Ginzburg-Landau equation for the charged condensates, while 
variation with respect to the vector potential defines 
Amp\`ere's law 
\Equation{EOM}{
   \D\D\psi_a=2\frac{\partial V(\Psi)}{\partial\psi_a^*}\, 
   ~~~~\text{and}~~~~
   \Curl\Curl\B=\J	\,,
}
with the supercurrent 
\Equation{Currents0}{
  \J\equiv\sum_{a}\J^{(a)}= 
   \sum_{a}\ea\Im\left(\psia^*\D\psia  \right)\,.
}
In the ground state, the condensates have constant densities 
$|\psi_a|=\sqrt{-\alpha_a/\beta_a}$ and $\A$ is a pure gauge. 
The length scales at which the condensates recover their ground 
state value after infinitesimal perturbations, the coherence lengths, 
are $\xi_a=1/\sqrt{-2\alpha_a}$. The penetration depth of the magnetic 
field $\lambda=1/e\sqrt{\sum_ag_a^2|\psi_a|^2}$ is consistently derived 
in section~\ref{LL}.

When considering vortex matter, we restrict ourselves to field 
configurations varying in the $xy$ plane only and with normal magnetic 
field, that is, field configurations describing both two-dimensional systems 
and three-dimensional system invariant under translations along the normal 
direction.
To investigate the physical properties of topological excitations 
in our model for mixtures of charged condensates, we numerically 
minimize the free energy \Eqref{FreeEnergy} within a finite element 
framework provided by the {\tt Freefem++} library \cite{Hecht:12}. 
For technical details, see the discussion in \Appref{Numerics}.

\section{Topological defects}
\label{Fractional}

Because we consider several condensates, the elementary topological 
excitations are fractional vortices, that is, field configurations 
with $2\pi$ phase winding of a single condensate (\eg $\varphi_1$ has 
$\oint\Grad\varphi_1=2\pi$ winding while $\oint\Grad\varphi_2=0$). 
A fractional vortex carries a fraction of the flux quantum. This can 
be seen by deriving the quantization condition for the magnetic flux. 
The supercurrent \Eqref{Currents0}, defined from Amp\`ere's 
equation $\Curl\B+\J=0$, reads as
\Equation{Currents}{
\J:=\frac{\delta\F}{\delta\A}=  
  e^2\varrho^2\A+e\sum_{a}\ga|\psi_a|^2\bs\nabla\varphi_a \,.
}
Here we defined the \emph{weighted} density  
$\varrho^2=\sum_a\ga^2|\psi_a|^2	$. Since the supercurrent $\J$ 
is screened, it decays exponentially and the magnetic flux thus 
reads as 
\Align{Flux}{
\Phi&=\int\B\cdot\bs{dS}=\oint \A \cdot\bs{d\ell} 	
	\nonumber\\
	&=\frac{1}{e^2\varrho^2}
	\oint \left(\J-e\sum_{a}\ga|\psi_a|^2\Grad\varphi_a\right) 
	\cdot\bs{d\ell}
	\nonumber \\
	&=-\frac{1}{e\sum_b\gb^2|\psib|^2}\sum_a\ga|\psia|^2
	\oint\Grad\varphi_a \cdot\bs{d\ell}\,.
}
Since the condensates $\psi_a$ are complex fields, their phases wind integer 
number of time. The couple $(k_1,k_2)$ denotes the field configurations 
with winding $k_a$ of the condensate $\psi_a$. The elementary excitations 
are fractional vortices $(1,0)$ and $(0,1)$ with unit winding in each 
component. A given fractional vortex in the condensate $a$ thus carries 
a fraction of the magnetic flux $\Phi_a/\Phi_0=\ga|\psia|^2/\varrho^2$. 
Here $\Phi_0=2\pi/e$ is the flux quantum. For the magnetic flux to be 
quantized, as long as $g_1\neq g_2$, it is necessary that different 
condensates have different winding number $k_a$. This follows from 
\Equation{Quantization}{
\sum_a\ga\frac{\Phi_a}{\Phi_0}=
	\frac{\sum_a\ga^2|\psia|^2}{\sum_b\gb^2|\psib|^2}=1 \,.
}
When both condensates carry the same electric charge, $g_1=g_2=1$, 
the quantization condition \Eqref{Quantization} reduces to 
the quantization conditions for multiband/multicomponent 
superconductors \cite{Babaev:02}.

\begin{figure*}[!htb]
\hbox to \linewidth{ \hss
\includegraphics[angle=90,width=\linewidth]{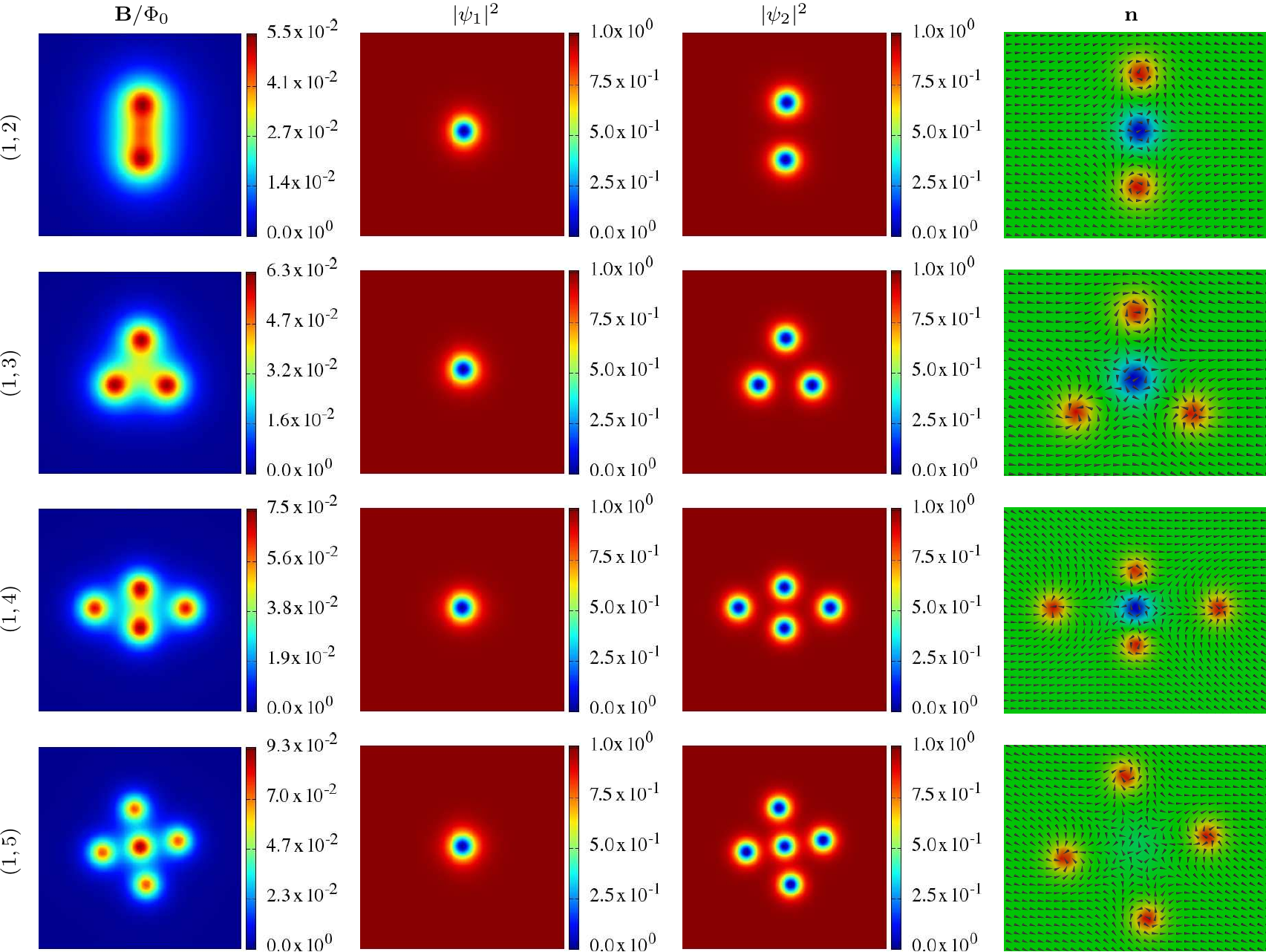}
\hss}
\caption{
(Color online) -- 
Molecule-like topological excitations carrying a unit flux quantum. 
The parameters of the Ginzburg-Landau functional \Eqref{FreeEnergy} 
are $(\alpha_a,\beta_a)=(-3,1)$, for both condensates ($a=1,2$) and 
$e=0.2$. Each row displays solutions for different winding parameters 
$(g_1,g_2)$ (indicated on left).
Displayed quantities in each row are respectively the magnetic field 
$\B$ (divided by the flux quantum $\Phi_0$) and the densities of both 
condensates $|\psi_1|^2$ and $|\psi_2|^2$ (in units of their ground 
state value). The rightmost panel shows the normalized projection 
of the pseudo-spin ${\bf n}$ \Eqref{Projection} onto the plane, while 
the color scheme indicates the magnitude of $n_z$. Blue corresponds to 
the south pole ($-1$) while red is the north pole ($+1$) of the target 
sphere $S^2$. 
}
\label{Fig:Vortex1}
\end{figure*}
Thus, each condensate $a$ has to wind $k_a=\ga$ times, so that the 
resulting composite vortex carries one flux quantum $\Phi_0=2\pi/e$. 
Fractional vortices have logarithmically divergent energy per unit length 
and thus cannot form in bulk systems \cite{Babaev:02}. This can be seen 
by rewriting the free energy into \emph{charged} and \emph{neutral} modes. 
For this, and using \Eqref{Currents}, the kinetic term can be rewritten 
\Align{Kinetic}{
\frac{1}{2}\sum_a|\D\psi_a|^2&=
	\frac{1}{2}\sum_a\Big\{(\Grad|\psi_a|)^2 
	+|\psi_a|^2(\Grad|\varphi_a|)^2  \Big\}	\nonumber \\
	&+\frac{\J^2}{2e^2\varrho^2} 
	-\frac{\left(\sum_ag_a|\psi_a|^2\Grad\varphi_a\right)^2}{2\varrho^2} \,,
}
with again the \emph{weighted} density $\varrho^2=\sum_a\ga^2|\psi_a|^2$. 
Defining the \emph{weighted} phase difference 
$\varphiot\equiv g_1\varphi_2-g_2\varphi_1$, the free energy 
\Eqref{FreeEnergy} reads as 
\SubAlign{GLRewritten1}{
\F&= \frac{1}{2}(\Curl \A)^2 + \frac{\J^2}{2e^2\varrho^2} 
   	\label{ChargedMode}\\
&+\sum_{a}\frac{1}{2}(\Grad|\psi_a|)^2
	+\alpha_a|\psi_a|^2+\frac{\beta_a}{2}|\psi_a|^4 
	\label{HiggsMode0} \\
&+\frac{|\psi_1|^2|\psi_2|^2}{2\varrho^2}(\Grad\varphiot)^2 
	\label{NeutralMode}	\,.
}
Since it decouples from the gauge field, the term \Eqref{NeutralMode} 
is called \emph{neutral mode}. This is the kinetic energy of the relative 
motion of the two condensates. That is, the co-directed (counter-directed) 
motion of particles with opposite (alike) charges $\ea$. 
When $\varphiot$ has a winding, this neutral mode has logarithmically 
divergent energy. Indeed, asymptotically each phase is well approximated 
by $\varphia=k_a\theta$, where $k_a$ are the (integer) vorticities and 
$\theta$ the polar angle. The condition for the logarithmic divergence 
to be absent and thus for the energy to be finite thus reads as
\Align{FiniteEnergy}{
0 & =\oint \Grad\varphiot \cdot\bs{d\ell} 
	=\oint \Grad(g_1\varphi_2-g_2\varphi_1) \cdot\bs{d\ell} \nonumber \\
  & = (g_1k_2-g_2k_1)  \,.
}
For a configuration carrying a single flux quantum (and since $\ga$ 
and $k_a$ are integers), the absence of winding in the weighted phase 
difference dictates that $k_a=g_a$. Thus the configurations which have 
no logarithmic divergence winds $g_1$ times in $\psi_1$ and $g_2$ 
times in $\psi_2$. Note that this condition also implies that the 
total flux is integer \Eqref{Quantization}. 
On the other hand, fractional vortices have logarithmically divergent 
energy. This is because they have winding in the \emph{weighted} phase 
difference. That is, when $g_2k_1\neq g_1k_2$, since the screening is 
incomplete, the energy of the vortex grows with the system size. 
Such vortices are typically thermodynamically unstable in bulk systems. 
When both condensates carry the same electric charge $g_1=g_2=1$, the 
condition \Eqref{FiniteEnergy} is automatically satisfied, provided both 
condensates have the same winding $k_1=k_2$. 
That is, only co-centered composite vortices with same winding in both 
condensates have finite energy. Since fractional vortices have logarithmically 
divergent energy per unit length, they cannot form in bulk systems 
\footnote{
A reservation should however be made here. 
Mesoscopic systems, because they introduce a boundary cut-off on the 
divergent mode,  can allow fractional vortices to be thermodynamically 
stable \cite{Chibotaru.Dao.ea:07,Chibotaru.Dao:10,Geurts.Milosevic.ea:08}.
Also fractional vortices can also be thermodynamically stable near 
boundaries \cite{Silaev:11}. In our paper we however only consider bulk
systems. In particular the simulation grid is chosen to be much larger 
than the size of vortices and thus boundary effects are irrelevant.
}
.

\begin{figure*}[!htb]
\hbox to \linewidth{ \hss
\includegraphics[width=\linewidth]{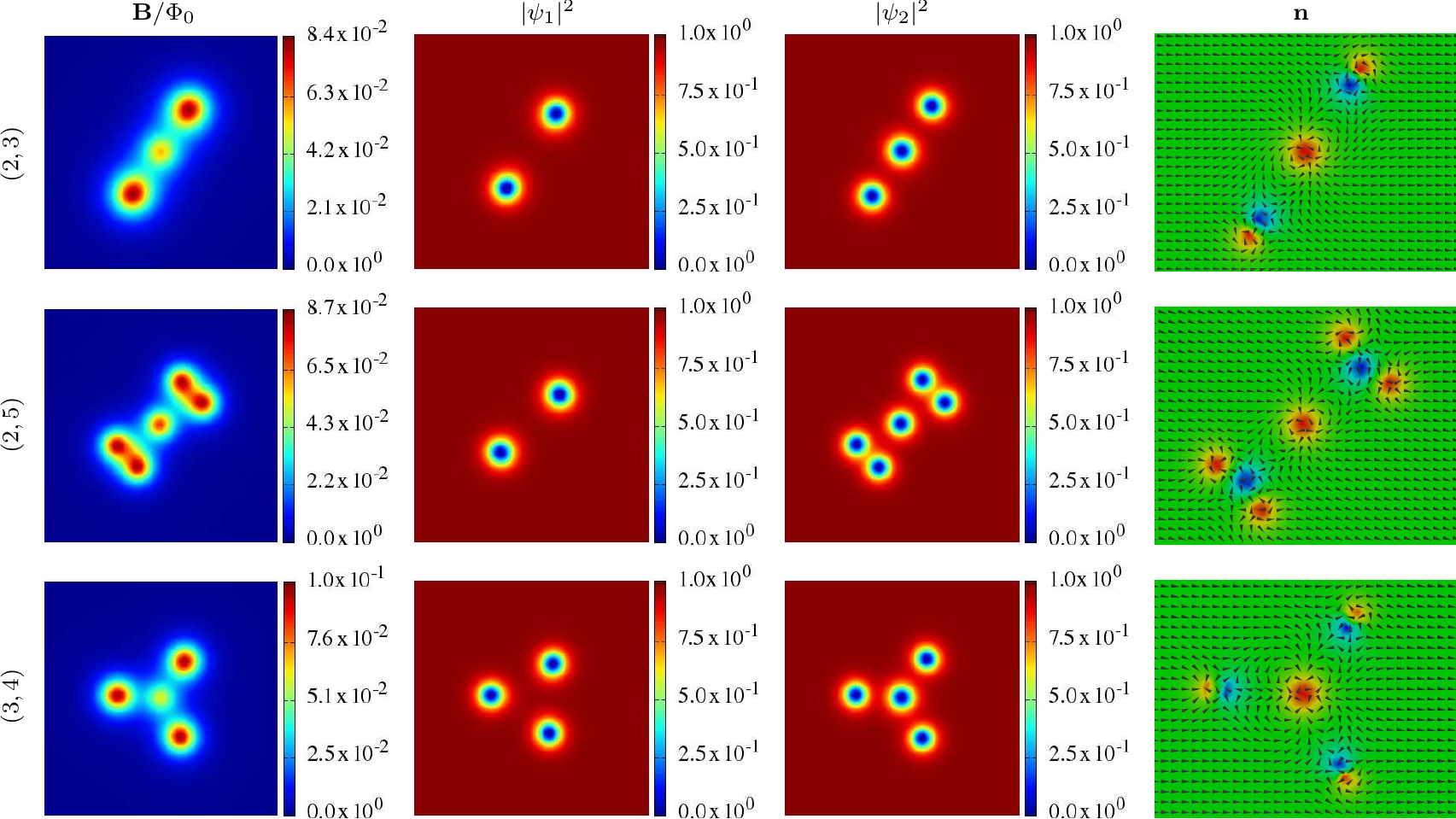}
\hss}
\caption{
(Color online) -- 
Molecule-like configurations of topological excitation carrying 
a unit flux quantum, for various values of winding parameters 
$(g_1,g_2)$ (indicated on the left).
The parameters of the Ginzburg-Landau functional and the displayed 
quantities are the same as in \Figref{Fig:Vortex1}.
}
\label{Fig:Vortex2}
\end{figure*}

Consider now the simplest case where the condensate $\psi_1$ carries 
single charge while $\psi_2$ carries double : $(g_1,g_2)=(1,2)$. The 
resulting composite vortex carrying one flux quantum is the bound state 
of one (fractional) vortex in $\psi_1$ and two vortices in $\psi_2$. To 
reduce the cost of kinetic energy in the neutral sector, the interaction 
through the neutral sector binds vortices together. It is minimal when 
cores in different condensates coincide (see discussion below in 
\Partref{LL}). However the two vortices in $\psi_2$ quite naturally repel 
each other, as would two Abrikosov vortices do in single band systems. 
As a result we can expect that if the magnetic repulsion is strong enough, 
despite the attractive channel through the neutral sector, fractional 
vortices will not overlap. 
This is indeed the case, as shown in \Figref{Fig:Vortex1}. The regime 
shown on the first line has $(g_1,g_2)=(1,2)$, so the condensates have 
single and double winding respectively. The resulting topological defect, 
carrying a single flux quantum, is a bound state of fractional vortices 
that are not co-centred and the configuration looks like an elongated rod. 
We refer to such bound state as a vortex molecule
\footnote{
Note that this physics of split vortices is entirely different from that 
of spatially separated fractional vortices that may also exist in different 
systems due to biquadratic density-density interaction \cite{Nitta.Eto.ea:14,
Agterberg.Babaev.ea:14,Kobayashi.Nitta:14a}, dissipationless drag interaction 
\cite{chung,Garaud.Sellin.ea:14} or coexistence of vortices and domain walls 
\cite{Garaud.Carlstrom.ea:13,Garaud.Carlstrom.ea:11}.
In the case of mixtures of condensates with commensurate charges, the 
splitting occurs because of competing attractive interaction through the 
neutral sector and repulsive interaction mediated by the magnetic field. 
As discussed below, unlike for core splitting induced by bi-quadratic terms, 
this physics is well captured in the London limit. 
}
. 

Bound states of non-overlapping fractional vortices feature special 
topological properties that motivate their denomination as skyrmions.
The terminology follows from the fact that two-component models 
can be mapped to easy-plane nonlinear $\sigma$-models 
that are associated with a \CPone topological invariant
\cite{Babaev.Faddeev.ea:02,Babaev:09,Garaud.Sellin.ea:14}. 
In systems with the same charge for both condensates, the two-component 
model is rewritten in term of the total current $\J$, the total 
density $\tilde{\varrho}^2=\sum_a|\psi_a|^2$, and the pseudo-spin $\bf n$. 
The pseudo-spin unit vector is the projection of the superconducting 
condensates on spin-$1/2$ Pauli matrices $\bs\sigma$: 
\Equation{Projection}{
 {\bf n}\equiv (n_x,n_y,n_z)
=\frac{\Psi^\dagger\bs \sigma\Psi}{\Psi^\dagger\Psi}\,.
}
When all condensates have the same couplings $\ga$, the spinor $\Psi$ 
is defined as the two-vector of complex condensates 
$\Psi^\dagger=(\psi_1^*,\psi_2^*)$. The finiteness of the energy dictates 
that $\bf n$ is asymptotically a constant vector, while vanishing of 
neutral modes implies that $n_x+in_y\propto\Exp{i(\varphi_2-\varphi_1)}$ 
has no winding.

To obtain a similar projection for incommensurate charges, the spinor 
$\Psi$ should be chosen so that the pseudo-spin does not wind asymptotically 
(when neutral mode vanish). There are several possibilities to realize 
such a projection and we choose 
\Equation{Spinor}{
\Psi^\dagger=\left(|\psi_1|\Exp{-ig_2\varphi_1},
|\psi_2|\Exp{-ig_1\varphi_2}\right) \,.
}
The projection \Eqref{Projection} of \Eqref{Spinor} maps to the 
two-sphere target space. It determines the pseudo-spin $\bf n$ that 
we use along the paper for the visualization of the pseudo-spin texture. 
Note that \Eqref{Spinor} is a non-holomorphic map, so it hard to justify 
the quantization of the associated invariant. For this we introduce another 
map which is holomorphic 
\Equation{Spinor2}{
\Psi^\dagger=\left(\psi_1^{*g_2},\psi_2^{*g_1}\right) \,.
}
The associated projection is a map from the one-point compactification of 
the plane ($\Real^2\cup\{\infty\}\simeq S^2 $) to the two-sphere target space 
spanned by $\bf n$. That is ${\bf n}: S^2\to S^2$, which is classified by 
the homotopy class $\pi_2(S^2)\in\Relative$. This defines the integer valued 
\CPone topological charge  
\Equation{Charge}{
   \Q({\bf n})=\frac{1}{4\pi} \int_{\Real^2}
   {\bf n}\cdot\partial_x {\bf n}\times \partial_y {\bf n}\,\,
  dxdy =g_1g_2\,.
}
If $\Psi\neq0$ everywhere, $\Q$ is an integer number. Roughly speaking, 
$\Q$ counts the number of times the texture $\bf n$ \Eqref{Projection} 
covers the target $S^2$ sphere. For practical purpose, and since it gives 
much better accuracy, we compute the degree of the maps \Eqref{Spinor} or 
\Eqref{Spinor2}, instead of computing the formula \Eqref{Charge}. 
Numerically calculated topological charge for the various configurations 
we constructed is indeed found to be integer (with a negligible error of 
order $10^{-4}$).

The map \Eqref{Spinor2} provides a rigorous justification of the topological 
invariant \Eqref{Charge}, but the associated texture $\bf n$ turns out to be 
very difficult to visualize. So we still use \Eqref{Spinor} that gives more 
straightforward physical interpretation for the visualization of the pseudo-spin 
texture. Interestingly, when computed numerically both definitions give 
similar result and accuracy of the topological charge.

In our parametrization of the gauge couplings $\ea=e\ga$, the charges 
of the condensates are commensurate. That is their ratio is a rational 
number and the simplest case we discussed is $(g_1,g_2)=(1,2)$. There, 
the unit flux quantum excitation is a molecule made of two fractional 
vortices in $\psi_2$ bound together by a single vortex in the $\psi_1$ 
condensate. For different ratio of the gauge couplings, the unit flux 
molecule-like bound states assume very rich structures as shown in 
\Figref{Fig:Vortex1} for $g_1=1$ and $g_2=2,3,4,5$. There, depending 
on the vorticities, fractional vortices can either be completely split 
apart or partially overlapping, as for example in $(g_1,g_2)=(1,5)$. 
The topological properties when fractional vortices overlap are essentially 
different than when they do not. Indeed the topological charge \Eqref{Charge}
is quantized only if there is no core overlap. To emphasize the richness 
in unit flux structures, we display in \Figref{Fig:Vortex2} configurations 
with other ratios of the commensurate charges. All of these have non-trivial, 
very different signature of the magnetic field. The structure of the 
molecule-like bound state depends not only on the ratio of the gauge 
couplings, but also on the ratio $m=|\psi_1|^2/|\psi_2|^2$ of the 
densities associated with each condensate. This can be seen from 
additional regimes we displayed in \Appref{App-material}.

\begin{figure*}[!htb]
\hbox to \linewidth{ \hss
\includegraphics[width=\linewidth]{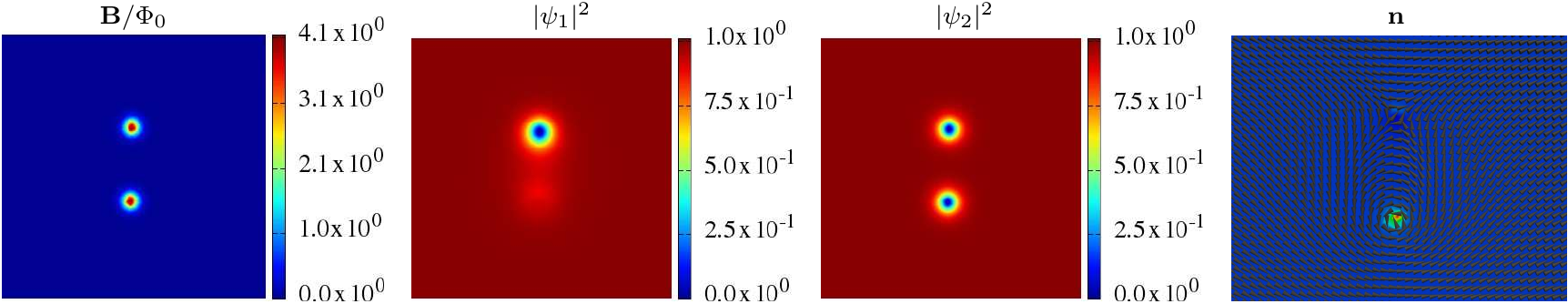}
\hss}
\caption{
(Color online) -- 
Vortex solutions carrying a single flux quantum for parameters of the 
Ginzburg-Landau functional \Eqref{FreeEnergy} giving a big disparity 
in condensate densities. The winding parameters here are $(g_1,g_2)=(1,2)$. 
The other parameters are $(\alpha_1,\beta_1)=(-10,10)$, 
$(\alpha_2,\beta_2)=(-10,0.01)$ and $e=0.2$. This choice follows 
crude estimation \Eqref{Estim} of relative condensate densities in 
the case of the superconducting state of liquid metallic deuterium. 
Displayed quantities are the same as in \Figref{Fig:Vortex1}. 
Note here that the pseudo-spin texture  $\bf n$, is mostly located 
\emph{almost} to the south pole of the target sphere because of the very 
big disparity in densities. That is, since $|\psi_1|^2\ll|\psi_2|^2$, 
then $n_z=\frac{|\psi_1|^2-|\psi_2|^2}{|\psi_1|^2+|\psi_2|^2}\approx-1$
everywhere except at the core of $\psi_2$ that does not overlap with 
the core in $\psi_1$. Note that since the core in $\psi_1$ coincides with 
a core in $\psi_2$, the south pole $n_z=-1$ is never reached.
}
\label{Fig:LMD}
\end{figure*}

As discussed later on in \Partref{LL}, depending on the ratio of densities 
in both condensates, the structure of the single flux quantum topological 
defect can be quite different. Indeed for substantial disparity in densities  
the `symmetric molecule' structure, dominated by long range quadrupolar 
mode of the relative phase $\varphi_{12}$, is no longer preferred. Instead, 
an `asymmetric molecule' is formed and it is characterized by a longer range 
dipolar mode of the relative phases. Such regimes with disparity in the 
condensate densities can be seen in \Figref{Fig:LMD}. More regimes are given 
as additional material in \Appref{App-material}.

\subsection*{A closer look to liquid metallic deuterium-like system}

\begin{figure}[!htb]
\hbox to \linewidth{ \hss
\includegraphics[width=\linewidth]{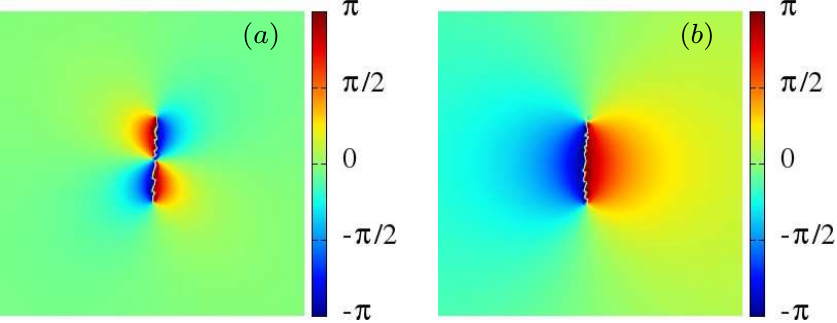}
\hss}
\caption{
(Color online) -- 
This shows the relative phases $\varphiot=g_1\varphi_2-g_2\varphi_1$, 
for single flux quantum topological defect with vorticities $(g_1,g_2)=(1,2)$.
The left panel $(a)$ with quadrupole structure corresponds to the regime 
in \Figref{Fig:Vortex1}. On the other hand, the right panel $(b)$ 
corresponding to \Figref{Fig:LMD} has dipole structure which is longer 
range. 
}
\label{Fig:PhaseDiff}
\end{figure}

We argued that our model captures topological aspects of mixtures of 
condensates with commensurate charges and that it qualitatively applies 
to various systems and for example to the superconducting state for liquid 
metallic deuterium (LMD), where one expects coexistence of electronic 
Cooper pairs and Bose condensate of deuterons.
Since reliable microscopic parameters for this state are not available 
we will study a phenomenological Ginzburg-Landau model for such a mixture 
of charged condensates.
The deuterium nucleus (deuteron) is a spin-1 particle that can condense in 
several states \cite{Oliva.Ashcroft:84,Oliva.Ashcroft:84a}. Here we ignore 
spin degrees of freedom and treat it as a scalar charged condensate carrying 
electric charge $+e$, while electronic Cooper pairs carry charge $-2e$. 
The mass of the electronic Cooper pairs is $m=2m_e\approx 1~\mathrm{MeV}/c$ 
while the mass of the deuteron is $m_d\approx 1875 ~\mathrm{MeV}/c$. Let $\psi_1$ 
and $\psi_2$ respectively denote the deuteronic and electronic condensates, 
so $(g_1,g_2)=(1,2)$.
Because of the electric neutrality at zero temperature we consider 
\Equation{Estim}{
m_1|\tilde{\psi}_1|^2=2m_2|\tilde{\psi}_2|^2\,,
}
(note for this expression we restore the $\tilde{~}$ symbols from 
\Eqref{Rescale}). Thus in this regime the electronic condensate is 
responsible for $99.9\%$ of the screening of the flux. 
In \Figref{Fig:LMD}, we show a single flux quantum topological defect 
for big disparity in ground state densities that are likely to occur 
for liquid metallic deuterium. There, the arrangement of fractional 
vortices is somewhat different from those displayed in \Figref{Fig:Vortex1}.
Indeed, unlike previously, some of the fractional vortices overlap.
The relative phase $\varphiot$ corresponding to the regime 
\Figref{Fig:Vortex1} assumes quadrupolar structure. The relative 
phases corresponding to \Figref{Fig:LMD}, instead shows a dipolar structure. 
This is shown in \Figref{Fig:PhaseDiff}. Dipole modes are longer range than 
quadrupole modes. This change in the long range behaviour of the relative 
phases should result in important modification of the large scale vortex 
matter structures. Long range dipolar modes were shown to play important 
role, although in a different context, on large scale vortex structure 
formation \cite{Garaud.Sellin.ea:14}. As discussed below in \Partref{LL}, 
the modification of the long range modes is consistently reproduced in the 
London approximation.

Here two constituent fractional vortices overlap. As a result, the 
topological invariant \Eqref{Charge} here is not quantized. This can be 
heuristically understood by the fact that the target sphere is not 
completely covered. With our crude estimates the ratio of condensate 
densities for liquid metallic deuterium, is about the same as for liquid 
metallic hydrogen. However because the commensurate charges are different, 
the topological excitations are completely different. The lowest energy 
topological excitation in liquid metallic hydrogen-like system is an 
axially-symmetric composite vortex, while for LMD it is a composite object 
of two co-centered vortices plus one satellite fractional vortex.  
The magnetic signature of the topological defect in LMD looks like a pair 
of vortices, while it is a single vortex for LMH.

\subsection*{The case of incommensurate charges}

The conditions for finite energy solutions \Eqref{FiniteEnergy} and the 
flux quantization \Eqref{Quantization} rely on the fact that charges are 
commensurate, that is, that their ratio is a rational number so that they 
can be parametrized as $\ea=e\ga$ where $\ga$ are integer numbers. 
For generality, here we address the question of what changes if charged 
condensates have incommensurate electric charges. When $\ea$ stands for 
elementary charges of elementary particles, they are integer multiples 
of an elementary electric charge. With current progress in creation of 
artificial gauge fields it could not be ruled out that systems with 
incommensurate coupling to the gauge field may be artificially realized. 

For this exercise, we have to relax the condition that $\ga$ are both 
integer numbers. Since $k_a$ have to be integer for the $\psi_a$'s 
to be single valued, the condition \Eqref{FiniteEnergy} ensuring both 
finite energy and flux quantization cannot be satisfied. As a result, 
the elementary vortices carry different flux $\Phi_a$ that cannot be 
added together to add up to a flux quantum. More precisely, 
the total flux \Eqref{Flux} is 
\Align{FluxBis}{
\Phi&= k_1\frac{2\pi g_1|\psi_1|^2/e}{g_1^2|\psi_1|^2+g_2^2|\psi_2|^2}
+k_2\frac{2\pi g_2|\psi_2|^2/e}{g_1^2|\psi_1|^2+g_2^2|\psi_2|^2} 
\nonumber \\
&=k_1\Phi_1+k_2\Phi_2\,,
}
and it is impossible to (consistently) write this as an integer times 
a flux quantum $\Phi_0$. 
Correspondingly, when charges are incommensurate, it is not possible 
by any mean, to eliminate the winding in the neutral sector 
\Eqref{NeutralMode}. Thus there are no finite energy solutions
(in infinite domain). Instead solutions have logarithmically divergent 
energy due to the (superfluid) mode associated with the neutral sector. 
That is, a phase gradient resulting from a phase winding always causes 
logarithmic divergence of vortex energy and cannot be fully compensated 
by the vector potential. 
 
Let us now close this remark about mixtures of condensates with 
incommensurate charges, and focus on the case where the ratio of 
the coupling constants $\ea$ is a rational number. That is 
$\ea=e\ga$ where $\ga$ are integers. We found that topological 
defects carrying integer flux are bound states of different 
fractional number of vortices in the different condensates. 
Because of the frustrated interactions that vortices in the same 
condensate repel, while they try to overlap with vortices in 
the other condensate, these bound states arrange into very 
complicated molecule-like structures. A large part of this 
exotic vortex structures can be captured by investigating the 
London limit, where fractional vortices can be mapped to Coulomb 
charges.

\section{London limit}
\label{LL}

When the electromagnetic repulsion is strong enough, integer vortices 
split to form a bound state of fractional vortices. The underlying 
physics describing the core splitting can be accurately captured within 
the London approximation where $|\psi_a|=\mathrm{const}$ everywhere 
(except for a sharp cut-off at vortex core). There, the expression 
\Eqref{GLRewritten1} further simplifies to
\SubAlign{GLRewritten2}{
   \F&= \frac{1}{2}\left(\B^2 + \frac{1}{e^2\varrho^2} 
   |\Curl\B|^2\right)\label{ChargedMode2}\\
 &+\frac{|\psi_1|^2|\psi_2|^2}{2\varrho^2}(\Grad\varphiot)^2 
\label{NeutralMode2}	\,.
}
The interaction energy of two non-overlapping fractional vortices 
is approximated in this London limit by considering \emph{charged} 
\Eqref{ChargedMode2} and \emph{neutral} modes \Eqref{NeutralMode2}, 
separately. The energy of the \emph{charged} sector \Eqref{ChargedMode2} 
reads as
\Equation{ChargedMode3}{ 
   F_{\mbox{\tiny mag}}=\int \frac{\B}{2}\left( 
\B+\lambda^2\Curl\Curl\B \right)\,,
}
where the London penetration length is $\lambda=1/e\varrho$.
The London equation for a (point-like) vortex placed at $\x_a$ 
and carrying a flux $\Phi_a$ is 
\Equation{Ampere}{
   \lambda^2\Curl\Curl\B+\B=\Phi_a\delta(\x-\x_a)\,,
}
and its solution is 
\Equation{Bfield}{
   \B_a(\x)=\frac{\Phi_a}{2\pi\lambda^2}
   K_0\left(\frac{|\x-\x_a|}{\lambda}\right)\,,
 }
where $K_0$ is the modified Bessel function of second kind. For two 
vortices located at $\x_a$ and $\x_b$, respectively carrying 
fluxes $\Phi_a$ and $\Phi_b$, the source term in the London equation 
reads as $\Phi_a\delta(\x-\x_a)+\Phi_b\delta(\x-\x_b)$ and the 
magnetic field is the superposition of two contributions 
$\B(\x)=\B_a(\x)+\B_b(\x)$. Thus  
\Align{Ampere2}{
    F_{\mbox{\tiny mag}}&=\int\frac{1}{2}
   (\B_a+\B_b )(\Phi_a\delta(\x-\x_a)+\Phi_b\delta(\x-\x_b)) \nonumber \\
   &=   \frac{\Phi_a\Phi_b}{2\pi\lambda^2}K_0
   \left(\frac{|\x_2-\x_1|}{\lambda}\right)
   +E_{va}+E_{vb} \,,
}
and $E_{va}\equiv\int\B_a(\x_a)\Phi_a/2$ is the (self-)energy 
of the vortex $a$. Finally, the interaction energy of two vortices 
in components $a,b$ reads as
\Equation{ChargedMode4}{ 
   E^{(int),\mbox{\tiny mag}}_{ab}=
   \frac{2\pi \ga\gb|\psi_a|^2|\psi_b|^2}{\varrho^2}
   K_0\left(\frac{|\x_a-\x_b|}{\lambda}\right)\,.
}
The interaction through the charged sector is thus screened interaction 
given by the modified Bessel function. When the couplings $e_a$ are 
parametrized such that they have the same sign (since the theory is invariant 
under complex conjugation, this is always possible), this interaction 
is always positive for any $a,b$ having the same sign of vorticity. It then 
gives, repulsive interaction between any kind of fractional vortices with 
co-directed winding. That is vortices repel while a vortex and an anti-vortex 
attract each other.
On the other hand, the interaction through the \emph{neutral} sector 
is attractive (\resp repulsive) for fractional vortices of the different 
(\resp same) condensate. The energy associated with the \emph{neutral} mode 
\Eqref{NeutralMode2} reads as 
\Equation{NeutralMode3}{ 
   F_{\mbox{\tiny neutral}}=
\frac{|\psi_1|^2|\psi_2|^2}{2\varrho^2}\int(\Grad\varphiot)^2\,.
}
To evaluate the interaction between fractional vortices in different 
condensates and respectively located at $\x_1$ and $\x_2$, the neutral 
sector is expanded 
\Align{NeutralModeLike1}{ 
   F_{\mbox{\tiny neutral}}=
\frac{|\psi_1|^2|\psi_2|^2}{2\varrho^2}\int&
(g_2\Grad\varphi_1)^2+(g_1\Grad\varphi_2)^2 \nonumber \\
&-2g_1g_2\Grad\varphi_1\cdot\Grad\varphi_2
\,.
}
At sufficiently large distance, a phase winding around some singularity 
located at a point $\x_a$, is well approximated by $\varphi_a=\theta$. 
Thus 
\Equation{GradPhi}{
\Grad\varphi_a=\frac{\Et}{|\x-\x_a|}={\bs z}\times\Grad\ln|\x-\x_a|\,.
}
As a result, the interaction part in the neutral sector reads as 
\Align{NeutralModeLike2}{ 
   E^{(int),\mbox{\tiny neutral}}_{12}&=
-\frac{g_1g_2|\psi_1|^2|\psi_2|^2}{\varrho^2}\int
	\Grad\varphi_1\cdot\Grad\varphi_2 \nonumber \\
&=2\pi g_1g_2\frac{|\psi_1|^2|\psi_2|^2}{\varrho^2}\ln|\x_2-\x_1|
\,.
}
Similarly, the interaction between two vortices in the same condensate 
$a$ is computed by requiring that the phase be the sum of the individual 
phases $\varphi_a=\varphi_a^\oo+\varphi_a^\ot$, while $\varphi_b=0$. 
Then the interaction reads as 
\Equation{NeutralModeUnLike}{ 
   E^{(int),\mbox{\tiny neutral}}_{aa}=
-2\pi\gb^2\frac{|\psi_1|^2|\psi_2|^2}{\varrho^2}\ln|\x_a^\ot-\x_a^\oo|
\,,
}
with here $b\neq a$.
To summarize, the interaction of vortices in different condensates is 
\Equation{InteractionUnlike}{
\frac{E^{(int)}_{12}}{2\pi}=g_1g_2\frac{|\psi_1|^2|\psi_2|^2}{\varrho^2}
    \Big(\ln \frac{r}{R} +wK_0\left(\frac{r}{\lambda}\right)\Big)\,,
}
while interactions of vortices of similar condensates are 
\Equation{InteractionLike}{
\frac{E^{(int)}_{aa}}{2\pi}= 
-\frac{\gb^2|\psi_1|^2|\psi_2|^2}{\varrho^2} \ln\frac{r}{R} 
+\frac{\ga^2|\psi_a|^4}{\varrho^2} K_0\left(\frac{r}{\lambda}\right)\,, 
}
with $b\neq a$, $r\equiv|\x_a-\x_b|$ and $R$ the sample size. Choosing 
the energy scale to be $2\pi g_1g_2|\psi_1|^2|\psi_2|^2/\varrho^2$  
and defining the parameters 
\Equation{Interaction_param}{
s = \frac{g_1}{g_2}\,,
~~~\text{and}~~~ m=\frac{|\psi_1|^2}{|\psi_2|^2} \,.	
}

\begin{figure}[!htb]
\hbox to \linewidth{ \hss
\includegraphics[width=\linewidth]{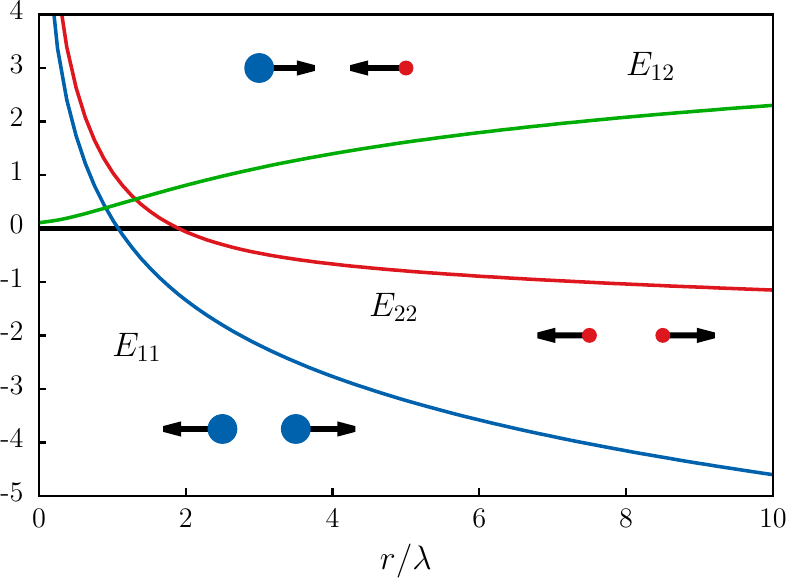}
\hss}
\caption{
(Color online) -- 
Interaction energies between point-like charges associated with 
vortices in different condensates. 
The blue (big) dot represents the vortex in $\psi_1$ while the 
red (small) dots represent the vortices in $\psi_2$.
Alike charges always repel while different charges attract.
Here we chose $m=0.2$ and $s=1/2$. 
Alike charges always repel while different charges attract with 
long range logarithmic attraction.
}
\label{Fig:Interaction}
\end{figure}
The interaction between fractional vortices reads as 
\Align{Interaction}{
E_{11}(r)&=\frac{1}{s}\ln\frac{R}{r}+\frac{m}{s}K_0\left(\frac{r}{\lambda}\right)
\,,\nonumber \\
E_{22}(r)&=s\ln\frac{R}{r}+\frac{s}{m}K_0\left(\frac{r}{\lambda}\right)	
\,,\nonumber \\
E_{12}(r)&=-\ln\frac{R}{r}+K_0\left(\frac{r}{\lambda}\right) \,.
}
Thus vortex matter in the London limit of a two-component superconductor 
with incommensurate charges is described by a 3-parameter family $(m,s,R)$. 
This is illustrated in \Figref{Fig:Interaction}.

The interaction between vortices in the same condensates is repulsive. 
In multi-component superconductors where both condensates have the same 
number of vortices in each component, vortices in different condensates 
will attract each other to form a bound state of co-centered vortices 
that minimizes the energy cost of the neutral sector \cite{Babaev:02,
Smiseth.Smorgrav.ea:05}. The situation here is more subtle. Because both 
condensates have different number of (fractional) vortices, the system 
has to compromise between vortices in similar condensates that repel each 
other and the fact that vortices in different condensates try to overlap. 
This explains why, beyond the London limit (see \eg \Figref{Fig:Vortex1}), 
integer vortices form a molecule-like bound state of split fractional vortices.

\subsection*{Transition in the structure of skyrmion}

\begin{figure}[!htb]
\hbox to \linewidth{ \hss
\includegraphics[width=\linewidth]{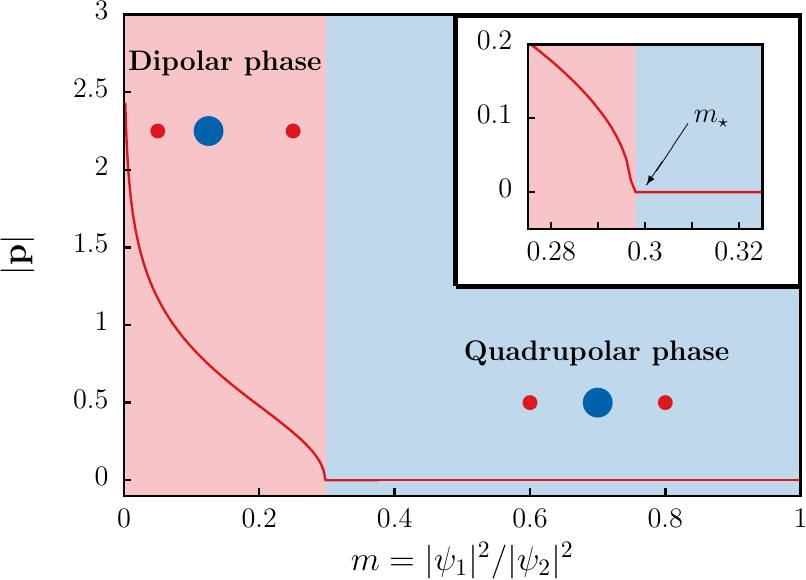}
\hss}
\caption{
(Color online) -- 
Structure of the single skyrmion for $g_1=1$ and $g_2=2$ as a 
function of relative ground state densities of the two condensates. 
There is a transition in the dipole moment of the molecule. The 
inset shows that transition closely. When ground state densities are 
quite similar, the molecule is symmetric and it has zero dipole moment. 
Below a certain threshold $m_\star$, the molecule becomes asymmetric 
and it develops a non zero dipole moment.
The blue (big) dot represents the vortex in $\psi_1$ while the 
red (small) dots represent the vortices in $\psi_2$
}
\label{Fig:Polarization}
\end{figure}

In the context of the mapping to point charges, the finite energy 
condition \Eqref{FiniteEnergy} is equivalent to require charge 
neutrality. Thus we complete the mapping to point charges by
defining the electric charge $q_2=g_1$ and $q_1=-g_2$. A neutral 
set of charged particle thus satisfy the charge neutrality
\Equation{ChargeNeutrality}{
	\sum_a q_ak_a=0\,.
}
We already know, from our solutions of the full non-linear model, 
that vortex solutions do exist and they have both finite energy 
and carry unit flux quanta. We also observed that, provided $e$ 
is small enough, that cores of fractional vortices are not superimposed. 
We now try to reproduce our results, using the equations 
\Eqref{Interaction} of the London limit. Here since a vortex configuration 
is a neutral set of discrete charge, it can be globally described 
by its dipole $p_i$ and quadrupole moments $d_{ij}$ in two dimensions 
\Align{Multipole}{
p_i=&\sum_aq_ar^\oa_i \,,	\nonumber\\
d_{ij}=&\sum_aq_a\left(2r^\oa_ir^\oa_j-\delta_{ij}r^{\oa\,2}_i\right)\,.
}

The minimum of the interaction energy \Eqref{Interaction} should 
describe the location of vortex cores. We apply this for a single 
vortex with $(g_1,g_2)=(1,2)$. Thus it is described by a set of 
three point particles, two carrying a single positive charge and 
one twice negatively charged. According to the interaction energies, 
it forms a bound state of non-overlapping particles, see 
\Figref{Fig:Polarization}.
Remarkably there is a transition of the `molecular' structure at 
a given $m=m_\star\approx 0.29714$. For $m>m_\star$, the molecule 
is symmetric and it has no dipole moment. The long-range interactions 
are thus dominated by quadrupole modes. For sufficient disparity in 
densities, when $m<m_\star$, the least energetic arrangement is no 
longer symmetric and thus the vortex molecule develops a dipole moment 
that is long range and should dramatically alter the large scale 
structures. Indeed, quadrupole modes are smaller in amplitude 
and also decay faster than dipole modes. For examples of the effect 
of long-range dipolar interactions in large scale structures 
(in a different context), see \cite{Garaud.Sellin.ea:14}. 

In our estimates to describe liquid metallic deuterium, the London 
limit parameter $m=|\psi_1|^2/|\psi_2|^2 \ll 1$. Thus according to 
the London limit picture \Figref{Fig:Polarization}, the single flux 
quantum vortex in the superconducting phase of liquid metallic deuterium 
should sit in the dipolar regime. As we illustrated in \Figref{Fig:LMD}, 
this is indeed the case, and there is perfect agreement between the 
London limit picture and direct numerical simulations.

Note that since the model describes independently conserved condensates, 
it has \groupUU invariance. Then both condensates have in principle 
different critical temperatures at which they condense. Then, there is 
in principle also always a regime where one of the condensates has much 
less density than the other. So there is always, at least, a regime 
where $m\ll1$ (\resp $m\gg1$) if $\psi_1$ (\resp $\psi_2$) condenses 
first. So there is always a phase dominated by the long range dipolar 
mode in the relative phases, that should dramatically influence large 
scale structuring of the vortex matter.

\subsection*{The case of many particles}

The physics of the topological defects carrying a single flux quantum 
is shown to be quite rich. Indeed even in the simplest case when 
$(g_1,g_2)=(1,2)$, the single skyrmion has a transition in its internal 
structure (see \Figref{Fig:Polarization}). The emerging long-range modes 
should have very important influence on the many-skyrmion states.
Investigating the many-skyrmion states, may give valuable information 
about the transport properties or about the lattice structures and their 
melting. Within the London limit, such properties can be investigated 
using molecular dynamics or Monte-Carlo simulations of the point particles 
interacting according to \Eqref{Interaction}. Although the point-charge 
model does not completely capture all the underlying physics, it can 
reproduce several aspects of the structures obtained beyond the London 
limit. This is beyond the scope of the current paper, yet we can address 
few general comments about the case of many particles.

In order to investigate the many-body properties of our model, one 
approach is to model fractional vortices by point charges with 
elementary interactions \Eqref{Interaction}, that is, to consider a 
set of $N_a$ particles corresponding to (fractional) vortices in the 
condensate $a$. For a system containing integer flux, the number of 
particles should satisfy the relation $g_2N_1=g_1N_2$. 
The interacting energy in the case of many particles reads as 
\Align{InteractionMany}{
E&= \sum_{i=1}^{N_1}\sum_{j>i}^{N_1}E_{11}
	\left(\left|\x^{(1)}_i-\x_j^{(1)}\right|\right) \nonumber \\
 &+ \sum_{i=1}^{N_2}\sum_{j>i}^{N_2}E_{22}
 	\left(\left|\x_i^{(2)}-\x_j^{(2)}\right|\right) \nonumber \\
 &+	\sum_{i=1}^{N_1}\sum_{j=1}^{N_2}E_{12}
 	\left(\left|\x_i^{(1)}-\x_j^{(2)}\right|\right)\,,
}
where $\x^{(a)}_i$ denotes the position of the $i$-th vortex of the 
condensate $a$ and interaction energies $E_{ab}$ are given by 
\Eqref{Interaction}. Note that this problem is related to the problem 
of unconventional plasma discussed in the context of quantum Hall states 
\cite{qhe1,qhe2,qhe3}.

The many-particle problem \Eqref{InteractionMany} can be investigated 
using different standard techniques such as molecular dynamics or 
Monte Carlo simulations. In the light of the complicated structure of 
the single skyrmions, one may expect very rich phases of the vortex matter 
there. However this deserves full investigation that is beyond the scope 
of the current paper.

\section{Conclusions}
\label{Conclusion}

We investigated physical properties of mixtures of charged (bosonic) 
condensates, carrying different electric charges. More precisely, 
we introduced a Ginzburg-Landau model that accounts phenomenologically 
for such mixtures. Disregarding the underlying microscopic theories 
that describe mixtures of charged condensates, this model is expected 
to qualitatively describe the topological excitations therein.

Elementary topological excitations are fractional vortices, that is, 
vortex configurations with winding in only one condensate. Because of 
the existence of a neutral mode, describing relative counter-directed 
motion of particles, fractional vortices have logarithmically divergent 
energy. 
The condition for having a finite energy solution in a mixture of 
condensates having commensurate electric charges $e_1=g_1e$ and $e_2=g_2e$ 
is that the phase should wind $g_1$ times in $\psi_1$ and $g_2$ times in 
$\psi_2$. Because of the commensuration of electric charge, finite-energy 
configurations have different number of fractional vortices in different 
condensates. 

Fractional vortices in the same condensate repel while fractional 
vortices in different condensates attract each other in order to 
reduce the energy cost associated with the counterflow of charge 
carriers. 
As a compromise, the topological excitation carrying an integer flux 
quantum can form a molecule-like bound state of fractional vortices, 
where there is no overlapping of vortices in contrast to systems  
with the same charges \cite{Babaev:02,Garaud.Sellin.ea:14}.

We also addressed the question of the underlying topology. There, 
we showed that two configurations carrying an integer flux quantum 
are differentiated from each other by a \CPone topological invariant. 
The topological excitations were explicitly constructed numerically 
and their structure, namely the spatial arrangement of constituent 
fractional vortices can be understood by investigating the London 
limit physics, where fractional vortices are mapped to point 
Coulomb charges.

The model we introduced and its topological excitations applies, 
at least qualitatively, to various physical systems where different 
condensates are formed and where they are commensurately coupled to 
the vector potential of the magnetic field. 
Namely it could effectively describe projected superconducting state 
of liquid metallic deuterium where deuterons form a charged Bose-Einstein 
condensate mixed with electronic Cooper pairs. This state of matter 
is currently a subject of experimental pursuit \cite{eremets}. Since 
these experiments are conducted in diamond anvil cell that can be equipped
with a receiving coil, the finding which we report could help to confirm 
or rule out formation of this state. 
Mixtures of commensurately charged condensates might also be an interesting 
system to be realized in cold atoms with synthetic gauge fields.

We acknowledge fruitful discussions with Johan Carlstr\"om, Karl Sellin, 
Daniel Weston and especially with J.M. Speight.
This work is supported by the Swedish Research Council, by the Knut 
and Alice Wallenberg Foundation through the Royal Swedish Academy of 
Sciences fellowship and by NSF CAREER Award No. DMR-0955902.
The computations were performed on resources provided by the Swedish 
National Infrastructure for Computing (SNIC) at National Supercomputer 
Center at Link\"oping, Sweden.

\appendix
\setcounter{section}{0}
\setcounter{equation}{0}
\renewcommand{\theequation}{\Alph{section}.\arabic{equation}}

\section{Additional material}
\label{App-material}

\begin{figure*}[!htb]
\hbox to \linewidth{ \hss
\includegraphics[angle=90,width=\linewidth]{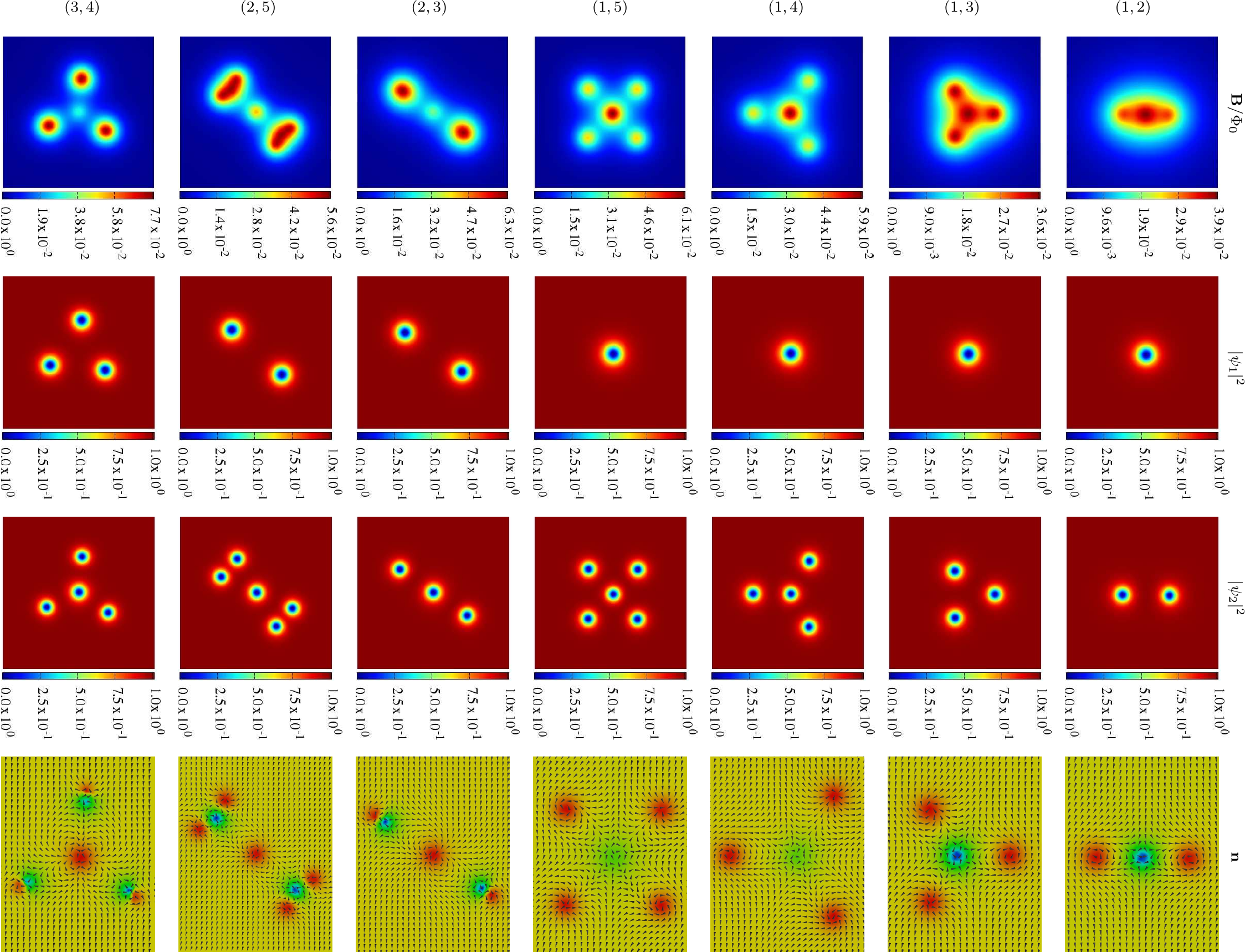}
\hss}
\caption{
(Color online) -- 
Single flux quantum topological excitations with disparity in the 
densities of each condensate. Here, the parameters of the Ginzburg-Landau 
functional \Eqref{FreeEnergy} are $(\alpha_1,\beta_1)=(-3,1)$, 
$(\alpha_2,\beta_2)=(-5,5)$ and $e=0.2$. Each row displays solutions 
for different winding parameters $(g_1,g_2)$ (indicated on left).
Displayed quantities in each row are respectively the magnetic 
field $\B$ (divided by the flux quantum $\Phi_0$) and the densities 
of both condensates $|\psi_1|^2$ and $|\psi_2|^2$ (in units of their 
ground state value). The rightmost panel displays the normalized 
projection of ${\bf n}$ onto the plane, while the color scheme indicates 
the magnitude of $n_z$. Blue correspond to the south pole ($-1$) while 
red is the north pole ($+1$) of the target sphere $S^2$. 
}
\label{Fig:AppVortex1}
\end{figure*}

\begin{figure*}[!htb]
\hbox to \linewidth{ \hss
\includegraphics[angle=90,width=\linewidth]{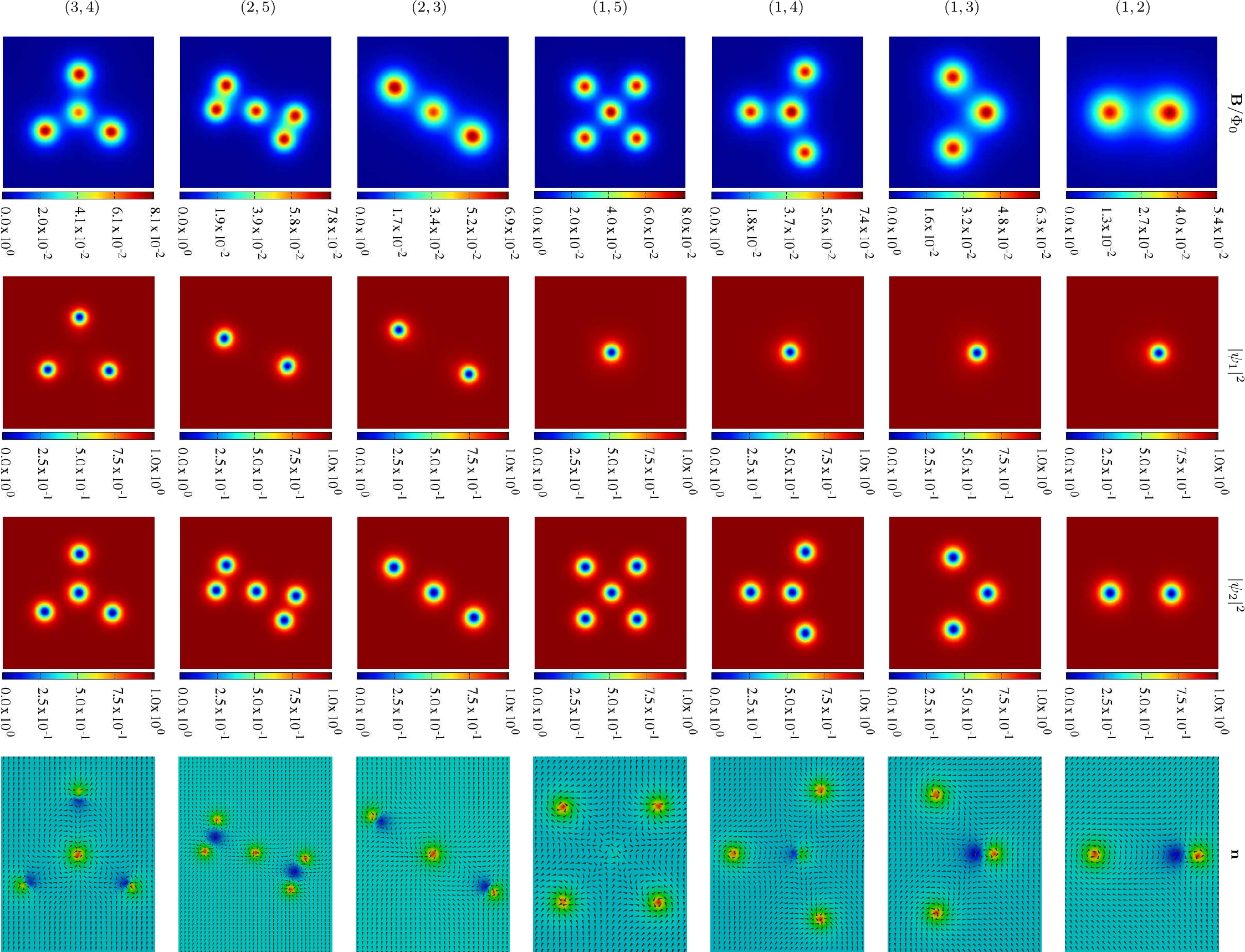}
\hss}
\caption{
(Color online) -- 
Vortex solutions carrying a single flux quantum. 
The parameters of the Ginzburg-Landau functional \Eqref{FreeEnergy} 
are $(\alpha_1,\beta_1)=(-5,5)$, $(\alpha_2,\beta_2)=(-3,1)$ and 
$e=0.2$. Each row displays solutions for different winding parameters 
$(g_1,g_2)$ (indicated on left).
Displayed quantities in each row are respectively the magnetic 
field $\B$ (divided by the flux quantum $\Phi_0$) and the densities 
of both condensates $|\psi_1|^2$ and $|\psi_2|^2$ (in units of their 
ground state value). The rightmost panel displays the normalized 
projection of ${\bf n}$ onto the plane, while color scheme indicates 
the magnitude of $n_z$. Blue correspond to the south pole ($-1$) while 
red is the north pole ($+1$) of the target sphere $S^2$. 
}
\label{Fig:AppVortex2}
\end{figure*}

In Figures \ref{Fig:AppVortex1} and \ref{Fig:AppVortex2}, we give 
additional single flux quantum skyrmions for different values of 
the parameters of the interacting potential \Eqref{Potential}.
In particular, we investigate here the role of $m=|\psi_1|^2/|\psi_2|^2$ 
parametrizing the relative ground-state densities of both charged 
condensates. 

\section{Finite element energy minimization}
\label{Numerics}

We consider the two-dimensional problem  \Eqref{FreeEnergy} 
defined on a domain $\Omega\subset\mathbbm{R}^2$ bounded by 
$\partial\Omega$. In our simulations, we choose the domain
$\Omega$ to be a disk. The problem is supplemented by the 
boundary condition ${\bs n}\cdot\D\psi_a=0$ with $\bs n$ the 
normal outgoing vector on $\partial\Omega$. This condition 
physically implies that no current flows through the boundary. 
This is thus a superconductor/insulator or superconductor/vacuum 
boundary condition.
Since this boundary condition is gauge invariant, additional 
constraint can be chosen on the boundary to fix the gauge. 
Our choice is to impose the radial gauge on the boundary 
$\bs e_\rho\cdot\A=0$ (note that with our choice of domain, 
this is equivalent to $\bs n\cdot\A=0$). This choice eliminates 
(most of) the gauge degrees and the boundary condition separates 
into two parts
\Equation{AppBC}{
   \bs n\cdot\Grad \psi_a=0~~~~~~\text{and}~~~~~~\bs n\cdot\A=0\,.
}
Note that these boundary conditions allow a topological defect 
to escape from the domain.
To prevent this in simulations of individual skyrmions or skyrmion
groups when no field is applied, the numerical grid is chosen to be 
large enough so that the attractive interaction with the boundaries 
is negligible. The size of the domain is then much larger than the 
typical interaction length scales. Thus, with this method one has to 
use large numerical grids, which is computationally demanding. 
This guarantees that the solutions are not boundary pressure artefacts.
In particular this means that the observed core splitting cannot 
be attributed to finite size effects as in mesoscopic samples 
\cite{Chibotaru.Dao.ea:07,Geurts.Milosevic.ea:08,Chibotaru.Dao:10}.

The variational problem is defined for numerical computation 
using a finite element formulation provided by the {\tt Freefem++} 
library \cite{Hecht:12}. Discretization within finite element 
formulation is done via a (homogeneous) triangulation over $\Omega$, 
based on Delaunay-Voronoi algorithm. Functions are decomposed 
on a continuous piecewise quadratic basis on each triangle. 
The accuracy of such method is controlled through the number of 
triangles, (we typically used $3\sim6\times10^4$), the order of 
expansion of the basis on each triangle (second order polynomial 
basis on each triangle), and also the order of the quadrature 
formula to compute the integral on the triangles. 

\subsection*{Initial guess}

The initial field configuration carrying $N$ flux quanta is prepared 
by using an ansatz which imposes phase windings around spatially 
separated $N_a=Ng_a$ vortices in each condensate: 
\Align{Initial_Guess1}{
\psi_a&= |\psi_a|\mathrm{e}^{ i\Theta_a+i\bar{\varphi}_a} 
\, ,~~ \nonumber \\
|\psi_a| &= u_a\prod_{k=1}^{N_a} 
\sqrt{\frac{1}{2} \left( 
1+\tanh\left(\frac{4}{\xi_a}({\cal R}^a_k(x,y)-\xi_a) 
\right)\right)}\, ,
}
where $a=1,2$, $u_a=\sqrt{-\alpha_a/\beta_a}$ is the ground-state 
density of a given condensate, and $\bar{\varphi}_a$ its ground-state 
phase. Because of \groupUU invariance, both $\bar{\varphi}_a$ 
can be chosen to be zero. $\xi_a$ parametrizes the size of cores while 
the functions
\Align{Initial_Guess2}{
\Theta_a(x,y)&=\sum_{k=1}^{N_a}
      \tan^{-1}\left(\frac{y-y^a_k}{x-x^a_k}\right)  \,,\nonumber\\
{\cal R}^a_k(x,y)&=\sqrt{(x-x^a_k)^2+(y-y^a_k)^2}\,. 
}
$(x^a_k,y^a_k)$ denotes the position of the singularity of the $k$-th 
vortex of the $a$ condensate. The starting configuration of the 
vector potential is determined by solving numerically Amp\`ere's 
equation on the background of the superconducting condensates given 
by Equations \Eqref{Initial_Guess1}--\Eqref{Initial_Guess2}. 

For a given starting configuration, the free energy is then minimized 
with respect to all degrees of freedom, with the condition \Eqref{AppBC} 
that no current flows through the boundary. Here we used a non-linear 
conjugate gradient method. The algorithm was iterated until relative 
variation of the norm of the gradient of the functional $\F$ with respect 
to all degrees of freedom was less than $10^{-6}$.

%

\end{document}